\documentclass{jfm}
\usepackage{color}
\usepackage{amstext}
\usepackage{amssymb}
\usepackage{graphicx}
\usepackage{amsmath}

\newcommand\const{\mathrm{const}}
\newcommand\Div{\mathrm{div}\,}

\newcommand\vX{\boldsymbol{X}}
\newcommand\vP{\boldsymbol{P}}

\newcommand\vV{\boldsymbol{V}}
\newcommand\va{\boldsymbol{a}}
\newcommand\vb{\boldsymbol{b}}

\newcommand\vf{\boldsymbol{f}}
\newcommand\vh{\boldsymbol{h}}

\newcommand\vK{\boldsymbol{K}}

\newcommand\vu{\boldsymbol{u}}
\newcommand\vv{\boldsymbol{v}}

\newcommand\vx{\boldsymbol{x}}

\newcommand\vp{\boldsymbol{p}}
\newcommand\vq{\boldsymbol{q}}
\newcommand\vQ{\boldsymbol{Q}}
\newcommand\vg{\boldsymbol{g}}

\newcommand\vxi{\boldsymbol{\xi}}

\newcommand\vkappa{\boldsymbol{\kappa}}

\renewcommand{\theequation}{\arabic{section}.\arabic{equation}}

\begin{document}

 {\title[Advection equation analysed by two-timing method] {Advection equation\\ analysed by two-timing method }}

\author[V. A. Vladimirov]{V.\ns A.\ns V\ls l\ls a\ls d\ls i\ls m\ls i\ls r\ls o\ls v}

\affiliation{DOMAS, Sultan Qaboos University, Oman and DAMTP, University of Cambridge, UK}

\pubyear{2016} \volume{xx} \pagerange{xx-xx}
\date{May 5, 2016}

\setcounter{page}{1}\maketitle \thispagestyle{empty}

\begin{abstract}

The aim of this paper is to study and classify the multiplicity of distinguished limits and asymptotic solutions for the advection equation with a general oscillating velocity field with the systematic use of the two-timing method. Our results are:

(i) the dimensionless advection equation contains \emph{two independent small parameters}, which represent the ratio of two characteristic time-scales and the spatial amplitudes of oscillations; the scaling of the variables and parameters contains Strouhal number;

(ii) an infinite sequence of \emph{distinguished limits} has been identified; this sequence corresponds to the successive degenerations of a \emph{drift velocity};

(iii) we have derived the averaged and oscillatory equations for the first \emph{four distinguished limits}; derivations are performed up to the forth orders in small parameters;

(v) we have shown, that \emph{each distinguish limit solution} generates an infinite number of \emph{parametric solutions}; these solutions differ from each other by the slow time-scale and the amplitude of prescribed velocity;

(vi) we have discovered the inevitable appearance of \emph{pseudo-diffusion}, which appears as a Lie derivative of the averaged tensor of quadratic displacements; we have clarified the meaning of pseudo-diffusion using a simple example;

(vii) our main methodological result is the introduction of a logical order into the area and classification of an infinite number of asymptotic solutions; we hope that it can help to study the similar problems  for more complex systems;

(viii) since in our calculations we
do not employ any additional assumptions, our study can be used as a test for the validity of the two-timing hypothesis;

(ix) the averaged equations for five different types of oscillating velocity fields have been considered as the examples of different drifts and pseudo-diffusion.

\end{abstract}

\section{Introduction}

This paper is devoted to the systematic studies of the asymptotic solutions of advection equation with  time-oscillating velocity field.
We employ the two-timing method in the form introduced by \cite{Vladimirov2, Yudovich,
Vladimirov1}, which allows us to produce the classification of an infinite number of distinguished limits and solutions.
We have obtained several new versions of averaged equations for a passive scalar and have presented several examples.
The main purpose of this paper is the introduction of a new systematic and general viewpoint on the hierarchy of distinguished limits, drifts, and pseudo-diffusion.
This viewpoint can help to organize and to unify a number of results, including that of \cite{Vladimirov2, Yudovich, VladimirovL, VladProc}.
We have chosen the advection equation as the simplest interesting example, with the hope that similar studies can be performed for other interesting equations.

The two-timing method has been used by many authors, see \cite{Kevorkian, Nayfeh, Verhulst}, however our analysis goes well beyond the usual calculations of the main approximations in various special cases.
Our analytic calculations are rigorous and straightforward by their nature, however they include a large number of integration by parts and algebraic transformations;
the performing of such calculations represents a `champion-type' result by itself.
The details of calculations are fully presented in the Appendix.

The key point of asymptotic  theory for ODEs or PDEs (with two or more small parameters) is the choice of asymptotic path in the space of these parameters. In other words, all available small parameters should be expressed through a single one; it allows one to compare different terms in asymptotic series. The related literature is very diverse and short of exact mathematical formulations.
Usually, during the studies of a particular problem, some `lucky' asymptotic path is chosen on the basis that it gives the successful derivation of an averaged equation or successful calculation of the main term in asymptotic solution, see \emph{e.g.} \cite{Kevorkian}.
Such `lucky' asymptotic paths are  called   \emph{distinguished limits} (DL's), while related solutions are called \emph{distinguished limit solutions} (DLS's).
Any studies of the multiplicity of distinguished paths (as well as the multiplicity of averaged equations and related solutions) and interrelations between them are always avoided.
A different approach is given by \cite{Yudovich}, where the concept of `strong',  `moderate', and `weak' oscillations is introduced. However, this approach overlooks some asymptotic solutions and could be seen as being `too rigid'; therefore we propose its development in this paper. Our asymptotic procedure is based on the simultaneous scaling (with the use of Strouhal number $St$) a slow time-scale and  vibrational velocity amplitude.
This procedure is flexible, self-consistent, covers broad classes of asymptotic solutions, and leads to a natural classification of solutions.

The considered advection equation is a hyperbolic PDE of the first order with time-oscillating coefficient, which represents  the \emph{prescribed} velocity field.
We emphasize that the velocity field has no connection to any particular dynamics.
Indeed, the governing equations for a velocity can be chosen very differently: they could represent an inviscid or viscous fluid, a fluid with any rheology, liquid crystals, blood, elastic and plastic medium, \emph{etc.} The advection equation can be also a part of the collisionless Boltzmann equation or Vlasov equation (see \emph{e.g.} \cite{Chierchia}).

The introduction of dimensional variables into the advection equation and the use of the two-timing assumption  leads us to the problem with \emph{two independent small scaling parameters that represent the ratio of two time-scales and  amplitude of oscillations}. The given set of parameters produces  the  \emph{Strouhal number $St$ as a basic large dimensionless parameter}; the appearance of $St$ means that the scaling is not unique. In agrement with the general \emph{ansatz} of asymptotic theory  we can use $St$ to vary all the characteristic scales; in our approach we use $St$ to introduce the appropriate scales of velocity and slow time.
The obtained dimensionless equation has been used for the identifying of distinguished limits. We have found
the infinite sequence of the distinguished limits, which are corresponding to the successive degeneration of a drift velocity.
The \emph{order} of this degeneration is chosen to enumerate the distinguished limits.
The derivation of averaged equations produce the successive approximations for a drift velocity as well as a qualitatively new `diffusion-like' term, which we call \emph{pseudo-diffusion}.
Remarkably, all the coefficients in the obtained averaged equations are universal; they are the same in DLS's of different orders and only `moving' to lower approximations with the increasing of the order of  DLS.
As the next important step  we discover \emph{one-parametric} families of solutions (infinite number of solutions corresponding to each DLS), different parametric solutions contain different slow time-scales and different velocity amplitudes.

Our results are particularly striking for the case of purely periodic velocity oscillations. Here we address an intriguing question: how to find the related slow time-scale while it does not appear in the coefficients of the equation? Our answer is:  the slow time-scale is uniquely determined by the magnitude of a given velocity in terms of $St$. For a given velocity we obtain the unique slow time-scale. We have established that the solutions corresponding to different orders of velocity are physically different even if they have the same functional form.

The central notion which appears in this paper is a \emph{drift velocity}. Our study can be seen as a  general view on the studies of  drift velocities with the emphasis at their different appearances in averaged equations; it is complemented by an unexpected but inevitable appearance of pseudo-diffusion.
We put the studies of drifts into the general context of the asymptotic theory, that has never been done before. In addition, there is a difference between the \emph{classical drift} and \emph{eulerian drift}. The classical (or lagrangian) drift appears as an averaged velocity for fixed lagrangian coordinates; hence the classical drift represents an average velocity of material particles.
The classical drift, in its main approximation, is a well-studied phenomenon, see  \cite{Stokes, Maxwell, Lamb, LH, Darwin, Batchelor, Lighthill, McIntyre, Craik, Grimshaw, Craik0, Hunt1, Eames, Buhler}. In our consideration (as well as in \cite{VladimirovL, VladProc}) the average operation takes place for \emph{fixed eulerian coordinates}, hence our drift represents an average velocity at a given eulerian position and we call it an \emph{eulerian drift}.
The relation between the classical drift and eulerian drift represents an open question to study, while their leading terms (in the amplitude series) are certainly coincide.
In our consideration of eulerian drift we explicitly show its different options, approximations, and appearances. We calculate first three  approximations for the eulerian drift velocity that is especially useful in the cases of vanishing leading approximation.

Situation with \emph{pseudo-diffusion} is more intriguing. First of all, the appearance of this `diffusion-like' term in averaged equations is surprising by itself.
Another surprise is the form of the pseudo-diffusion matrix which appears as a \emph{Lie-derivative} of an averaged \emph{quadratic displacement tensor}.
For different flows, pseudo-diffusion can correspond to diffusion or `anti-diffusion' or to some intermediate cases.
We make the first step in the understanding of  the meaning of pseudo-diffusion by considering  some simple example of rigid-body oscillations. We arrive to a surprising conclusion  that the combination of the advection due to rigid-body oscillations and the eulerian averaging procedure can produce the evolution of  the averaged field, which formally coincides with physical diffusion.

We present five examples of flows aimed to illustrate different appearances of drift and pseudo-diffusion. These examples cover the general form of modulated oscillatory fields, Stokes drift, a spherical `acoustic' wave, a class of velocity fields with vanishing main term of the drift, and a velocity field which provides a chaotic averaged `lagrangian' dynamics.

It is important, that in this paper we do not use any additional assumptions, except the two-timing hypothesis itself. All the results appear from the straightforward and rigorous calculations and represent mathematical theorems, although they are not formulated explicitly in this way.
Hence, our results can be seen as the test on the validity and reliability of two-timing method. Any failure to interpret our results physically or mathematically can be explained only by the insufficiency of two time-scales, that might mean that three time-scales (or more) are required. At the same time, following to a high mathematical rigour, we avoid any physical interpretations, while such interpretations could be needed in future.

\section{Formulation of problem and two-timing assumption}

A fluid flow is given by its velocity field $\vv^*(\vx^*,t^*)$, where $\vx^*=(x_1^*,x_2^*,x_3^*)$ and $t^*$ are cartesian coordinates and time, asterisks stand for dimensional variables.
We suppose that velocity field is sufficiently smooth, but \emph{we do not assume that it satisfies any equations of motion}.
The dimensional advection equation for a scalar field $a(\vx^*,t^*)$ is
\begin{eqnarray}
a_{t^*}+(\vv^*\cdot\nabla^*) a=0,\qquad a_{t^*}\equiv \partial a/\partial {t^*}
\label{exact-1}
\end{eqnarray}
This equation describes the motions of
a lagrangian marker in either an incompressible or compressible fluid, and the advection of a passive scalar admixture
with concentration $a(\vx^*,t^*)$ in an incompressible fluid. It is also closely relevant to various models, not related to fluid dynamics.
The hyperbolic equation (\ref{exact-1}) has characteristics curves (trajectories) $\vx^*=\vx^*(t^*)$ described by an ODE
\begin{eqnarray}
{d\vx^*}/{dt^*}=\vv^*(\vx^*,t^*),\qquad \vx^* |_{t=0}=\vX^*
\label{exact-1a}
\end{eqnarray}
where $\vx^*$ and $\vX^*$ are eulerian and lagrangian coordinates.
The \emph{classical  drift} motion follows after the averaging of (\ref{exact-1a}), which is automatically performed for fixed lagrangian coordinates.
In contrast, in this paper we consider only the equation (\ref{exact-1}), which is subject of averaging for fixed eulerian coordinates, the related drift motion can be called \emph{eulerian drift.}
Such an approach significantly simplifies calculations and is reacher on results, however the link between the classical (lagrangian) drift and the considered below eulerian drift emerges as a new problem to study.

 We accept that the field $\vv^*(\vx^*,t^*)$ is oscillating in time and possesses the characteristic scales of velocity $U$, length $L$, and frequency $\omega^*$.
These three parameters give a dimensionless parameter -- Strouhal number $St\equiv L\omega^*/U$, hence the dimensionless variables and parameters (written below without asterisks) \emph{are not unique}; in this paper we use the following set
\begin{eqnarray}
&& \vx^*= L\vx,\ t^*=(L/U)t,\ \omega^*=(U/L)\omega,\ \vv^*=U(St)^\beta\vv\label{variables}
\end{eqnarray}
while  $\vv$ has the `two-timing' functional form
\begin{eqnarray}
&& \vv =\vu(\vx, s, \tau),\  \tau\equiv\omega t,\quad s\equiv t/\omega^\alpha;\quad \vu\sim O(1)\label{variables1}
\end{eqnarray}
with constants
\begin{eqnarray}
&& \alpha=\const >-1,\ \beta=\const<1,\ \omega=St\gg 1 \label{exact-2a}
\end{eqnarray}
We note that the scale of velocity is $U(St)^\beta$, not just $U$; such kind of `combined' scaling is usual for asymptotic analysis and for  fluid dynamics (for example, for viscous flows the length-scale can be chosen as $L\sqrt{Re}$, where $Re$ is Reynolds number).
In fact, the accepted scaling for a velocity is absolutely required for the further consideration, otherwise (say, for $\vv^*=U\vv$) the asymptotic solutions for some important cases do not exist (see Discussion).
The appearance of two constants $\alpha$ and $\beta$ reflects the fact that we have two small parameters in the equation.
Different options for the values of  $\alpha$ and $\beta$ will be introduced later.
Physically, the restriction $\alpha>-1$  makes the variable $s$ `slow' in comparison with $\tau$,  while $\beta<1$ gives the smallness of a vibrational spatial amplitude.

In the accepted dimensionless  variables (after the use of the chain rule) our main equation (\ref{exact-1}) takes the form
\begin{eqnarray}
&&\omega a_\tau+ \frac{1}{\omega^\alpha} a_s+\omega^\beta({\vu}\cdot\nabla)a= 0, \qquad {\partial}/{\partial{t}}=\omega\partial/\partial\tau+\frac{1}{\omega^\alpha}\partial/\partial s\label{exact-6}
\end{eqnarray}
where subscripts $\tau$ and $s$  stand for the related partial derivatives; $s$ and $\tau$ represent two \emph{mutually dependent} variables, which are called \emph{slow time} and \emph{fast time}.
Equation (\ref{exact-6}) can be rewritten in the form containing two independent small parameters
\begin{eqnarray}
&&a_\tau+\varepsilon_2 a_{s}+\varepsilon_1 \vu\cdot\nabla a= 0;\quad \varepsilon_1\equiv \omega^{(\beta-1)}, \quad \varepsilon_2\equiv 1/\omega^{(\alpha+1)}
\label{exact-6a}
\end{eqnarray}
Hence,  we study the asymptotic limit in the plane $(\varepsilon_1, \varepsilon_1)$
\begin{eqnarray}
&&(\varepsilon_1,\varepsilon_2)\to (0,0)
\label{paths}
\end{eqnarray}
where different asymptotic paths can be prescribed by different choices of $\alpha$ and $\beta$ in \eqref{exact-6a}.
Each such a path may produce different solutions; the paths which produce valid asymptotic solutions are called \emph{distinguished limits}.

\section{Suggestions and notations}

The key suggestion of the two-timing method is
\begin{eqnarray}
&&\emph{$\tau$ and $s$ are considered as mutually independent variables}\label{key}
\end{eqnarray}
As a result, we convert \eqref{exact-6}, \eqref{exact-6a} from PDE with independent variables $t$ and $\vx$  (where $\tau$ and $s$ are mutually dependent   \eqref{variables1} and expressed in terms of $t$) into  PDE with the extended number of independent variables $\tau, s$ and $\vx$. Then the solutions of \eqref{exact-6}, \eqref{exact-6a} must have a functional form:
\begin{eqnarray}
&& a=a(\vx, s, \tau)\label{exact-3}
\end{eqnarray}
It should be emphasized, that without \eqref{key} the functional form of solutions can be different from \eqref{exact-3}; indeed the presence of $St$ allows us to build an infinite number of different time-scales, not just $\tau$ and $s$. In this paper we accept \eqref{key} and analyse the related averaged equations and solutions in the functional form \eqref{exact-3}.

In order to make further analytical progress, we introduce a few convenient notations and agreements.
In this paper we assume that \emph{any dimensionless function} $f(\vx,s,\tau)$ has the following properties:

(i)  $f\sim {O}(1)$ and  all its $\vx$-, $s$-, and $\tau$-derivatives of the required below orders are also ${O}(1)$;

(ii) $f$  is $2\pi$-periodic in $\tau$, \emph{i.e.}\ $f(\vx, s, \tau)=f(\vx,s,\tau+2\pi)$ (about this technical simplification see Discussion);

(iii) $f$ has an average given by
\begin{equation}\label{average}
  \langle {f}\,\rangle \equiv \frac{1}{2\pi}\int_{\tau_0}^{\tau_0+2\pi}
f(\vx, s, \tau)\, d \tau \qquad \forall\ \tau_0=\const;
\end{equation}

(iv) $f$ can be split into averaged and purely oscillating parts
\begin{equation}\label{split}
  f(\vx, s, \tau)=\overline{f}(\vx, s)+\widetilde{f}(\vx, s, \tau)
\end{equation}
where  \emph{tilde-functions} (or  purely oscillating functions) are such that $\langle \widetilde f\, \rangle =0$ and the \emph{bar-functions} $\overline{f}(\vx,s)$ (or the averaged functions $\langle {f}\,\rangle=\overline{f}(\vx,s)$) are $\tau$-independent;

(v)  we introduce a special notation $\widetilde{f}^{\tau}$ (with a superscript $\tau$) for  the
\emph{tilde-integration} of tilde-functions, such integration keeps the result in the tilde-class. We notice that the integral of a tilde-function
\begin{equation}
F(\vx,s,\tau)\equiv\int_0^\tau \widetilde{f}(\vx,s,\tau')\, d \tau'
\label{ti-integr0}
\end{equation}
often does not belong to the tilde-class. In order to keep the result of integration in the tilde-class we should subtract the average $\widetilde{F}=F-\overline{F}\equiv \widetilde{f}^\tau$, which can be rewritten as
\begin{equation}
\widetilde{f}^{\tau}\equiv\int_0^\tau \widetilde{f}(\vx,s,\tau')\, d \tau'
-\frac{1}{2\pi}\int_0^{2\pi}\Bigl(\int^\mu_0
\widetilde{f}(\vx,s,\tau')\, d \tau'\Bigr)d\mu.
\label{ti-integr}
\end{equation}

\section{Distinguished limits and averaged equations}
 Our calculations have shown that there is a series of \emph{distinguished limits} for the equation (\ref{exact-6a}) with independent variables $\tau, s, \vx$; each distinguished limit represents the one-parametric path $\varepsilon_1=\varepsilon$, $\varepsilon_2=\varepsilon^n$ (with an integer $n>0$) in the plane $(\varepsilon_1,\varepsilon_2)$; hence (\ref{exact-6a}) yields:
\begin{eqnarray}
&&a_\tau+\varepsilon^n a_{s}+\varepsilon \vu\cdot\nabla a= 0
\label{exact-6c}
\end{eqnarray}
  We denote the distinguished limits as DL($n$); our calculations have been performed for $n=1,2,3,4$. The detailed calculations for the most interesting case $n=2$ are given in Appendix. Other cases $n=1,3,4$ are very similar and contain the same coefficients of the averaged equations and the same blocks of calculations as the case $n=2$; therefore for $n=1,3,4$ we present below some final results only.
In all cases we are looking for solutions in the form of regular series
\begin{eqnarray}
a(\vx,s,\tau)=\sum_{k=0}^\infty\varepsilon^k a_k(\vx,s,\tau),\quad a_k=\overline{a}_k(\vx,s)+\widetilde{a}_k(\vx,s,\tau),\quad k=0,1,2,\dots
\label{basic-4aa}
\end{eqnarray}
Analytical calculations contain the following two steps:
(i) writing  the equations of successive approximations and splitting each such equation into its `bar' and `tilde' parts, see \eqref{split};\ (ii) obtaining the closed systems of equations for the `bar' parts; during this step one should  perform a large number of integrations by parts and algebraic transformations.  The results give  full solutions $a=\overline{a}+\widetilde{a}$ in any considered approximation. The described above steps can be performed only for distinguished limits; all other asymptotic paths \eqref{paths} produce controversial systems of equations or they are not leading to closed systems of equations for successive approximations.
The first four DL's correspond to the variables $s$ and velocity $\vu$ as following:

DL(1):  $s= t$;  $\vu=\overline{\vu}(\vx,s)+\widetilde{\vu}(\vx,s,\tau)$, where both `bar' and `tilde' parts of $\vu$ are not zero;

DL(2): $s=\varepsilon t$;\  $\vu=\widetilde{\vu}(\vx,s,\tau)\neq 0$, where $\overline{\vu}(s,\vx)\equiv 0$;

DL(3):  $s=\varepsilon^2 t$;\  $\vu=\widetilde{\vu}(\vx,s,\tau)$,  where $\overline{\vu}(s,\vx)\equiv 0$, and $\overline{\vV}_0\equiv 0$;

DL(4): $s=\varepsilon^3 t$;\  $\vu=\widetilde{\vu}(\vx,s,\tau)$, where $\overline{\vu}(s,\vx)\equiv 0$, $\overline{\vV}_0\equiv 0$, and $\overline{\vV}_1\equiv 0$.

\noindent The accepted notations are
\begin{eqnarray}
&&\vV_0\equiv \frac{1}{2}[\widetilde{\vu},\widetilde{\vxi}],\quad
\overline{\vV}_1\equiv\frac{1}{3}\langle[[\widetilde{\vu},\widetilde{\vxi}],\widetilde{\vxi}]\rangle,\quad \widetilde{\vxi}\equiv\widetilde{\vu}^\tau;
\label{4.18}
\end{eqnarray}
where the square brackets stand for a commutator of any two vector-functions $\vf$ and $\vg$
 $$
 [\vf,\vg]\equiv(\vg\cdot\nabla)\vf-(\vf\cdot\nabla)\vg
 $$
From the above list, one can see that the increasing of $n$ corresponds to the successive degenerations of velocity $\vu$ and drift velocity  $\overline{\vV}\equiv 0$.

The case DL(1) with $\overline{\vu}=O(1)\neq 0$ corresponds to the advection speed of order one, hence $s=t$ and the averaged equation of zero approximation is
\begin{eqnarray}\nonumber
\left(\partial_s+ \overline{\vu}\cdot\nabla\right)\overline{a}_{0}=0
\end{eqnarray}
One can see that in the main approximation we have the advection of an averaged scalar field with the averaged velocity. However, in this paper we concentrate our attention to the cases when the oscillatory part of a velocity enters the averaged equations of the main approximation. Therefore, we do not consider DL(1) in detail here, we just state that all the coefficients in the averaged equations are similar to that in DL(2) (see below); more exact: the same coefficients appear in the DL(1) averaged equations of the next approximation in $\varepsilon$ (in comparison with DL(2)).

DL(2) is the most instructive and physically interesting case, therefore we consider it in detail.
Here, we have $\overline{\vu}\equiv 0$, hence, the speed of
advection is described by a longer (than in DL(1)) slow time-scale $s=\varepsilon t$.
The averaged equations of the first three successive approximations are (see the detailed derivation in Appendix)
\begin{eqnarray}
&&\left(\partial_s+ \overline{\vV}_0\cdot\nabla\right)\overline{a}_{0}=0,\quad \partial_s\equiv \partial/\partial s\label{4.15}\\
&&\left(\partial_s+ \overline{\vV}_0\cdot\nabla\right)
\overline{a}_{1}+(\overline{\vV}_1\cdot\nabla)\overline{a}_0=0\label{4.16}\\
&&\left(\partial_s+ \overline{\vV}_0\cdot\nabla\right)\overline{a}_{2}+
(\overline{\vV}_1\cdot\nabla)\overline{a}_1+(\overline{\vV}_2\cdot\nabla)\overline{a}_0=
\frac{\partial}{\partial x_i}\left(
\overline{\chi}_{ik}\frac{\partial\overline{a}_0}{\partial x_k}\right),
\label{4.17}
\end{eqnarray}
with notations
\begin{eqnarray}
&&\overline{\vV}_2\equiv\frac{1}{4}\langle[[\vV_0,\widetilde{\vxi}],\widetilde{\vxi}]\rangle +
\frac{1}{2}\langle[\widetilde{\vV}_0,\widetilde{\vV}_0^\tau]\rangle+
\frac{1}{2}\langle[\widetilde{\vxi},\widetilde{\vxi}_s]\rangle
+\frac{1}{2}\langle\widetilde{\vxi}\Div\widetilde{\vu}'+ \widetilde{\vu}'\Div\widetilde{\vxi}\rangle,\label{4.19}
\\
&&\widetilde{\vu}'\equiv\widetilde{\vxi}_s-[\overline{\vV}_0,\widetilde{\vxi}],
\label{4.20a}\\
&&2\overline{\chi}_{ik}\equiv\langle\widetilde{u'}_i\widetilde{\xi}_k+\widetilde{u'}_k\widetilde{\xi}_i\rangle=
\mathfrak{L}_{\overline{\vV}_0}\langle\widetilde{\xi}_i\widetilde{\xi}_k\rangle,\label{4.20}\\
&&\mathfrak{L}_{\overline{\vV}_0}\overline{f}_{ik}\equiv
\left(\partial_s+
\overline{\vV}_0\cdot\nabla\right)\overline{f}_{ik}-\frac{\partial\overline{V}_{0k}}{\partial x_m}\overline{f}_{im}-
\frac{\partial\overline{V}_{0i}}{\partial x_m}\overline{f}_{km}
\label{4.20b}
\end{eqnarray}
where the operator $\mathfrak{L}_{\overline{\vV}_0}$ is such that $\mathfrak{L}_{\overline{\vV}_0}
\overline{f}_{ik}=0$ represents the condition for tensorial field $\overline{f}_{ik}(\vx,t)$
to be `frozen' into $\overline{\vV}_0(\vx,t)$ ($\mathfrak{L}_{\overline{\vV}_0}$ is also known as the \emph{Lie derivative} of a vector field).
The summation convention is in use everywhere in this paper.

Three equations (\ref{4.15})-(\ref{4.17}) can be written
as a single advection-`diffusion' equation (with an error ${O}(\varepsilon^3)$)
\begin{eqnarray}
&&\left(\partial_s+ \overline{\vV}\cdot\nabla\right)\overline{a} =
\frac{\partial}{\partial x_i}\left(\overline{\kappa}_{ik}\frac{\partial\overline{a}}{\partial x_k}\right)
\label{4.21}\\
&&\overline{\vV}=\overline{\vV}_0+\varepsilon\overline{\vV}_1+\varepsilon^2
\overline{\vV}_2,\label{4.22}\\
&&\overline{\kappa}_{ik}=\varepsilon^2\overline{\chi}_{ik}\label{4.23}\\
&&\overline{a}=\overline{a}_0+\varepsilon\overline{a}_1+\varepsilon^2\overline{a}_2
\label{4.22aa}
\end{eqnarray}
Eqn. (\ref{4.21}) shows that the averaged motion represents a drift with velocity $\overline{\vV}+{O}(\varepsilon^3)$ and
\emph{pseudo-diffusion } with matrix coefficients
$\overline{\kappa}_{ik}+{O}(\varepsilon^3)$.

For DL(3) we impose a restriction $\overline{\vV}_0\equiv 0$ and (similarly to the calculations in Appendix) derive the equations:
\begin{eqnarray}
&&\left(\partial_s+ \overline{\vV}_1\cdot\nabla\right)\overline{a}_{0}=0\label{4.15a}\\
&&\left(\partial_s+ \overline{\vV}_1\cdot\nabla\right)\overline{a}_{1}+
(\overline{\vV}_2\cdot\nabla)\overline{a}_0=
\frac{\partial}{\partial x_i}\left(
\overline{\chi}_{ik}\frac{\partial\overline{a}_0}{\partial x_k}\right),
\label{4.17a}\\
&&\overline{\vV}_2\equiv\frac{1}{4}\langle[[\widetilde{{\vV}}_0,\widetilde{\vxi}],\widetilde{\vxi}]\rangle +
\frac{1}{2}\langle[\widetilde{\vV}_0,\widetilde{\vV}_0^\tau]\rangle+
\frac{1}{2}\langle[\widetilde{\vxi},\widetilde{\vxi}_s]\rangle
+\frac{1}{2}\partial_s\langle\widetilde{\vxi}\Div\widetilde{\vxi}\rangle,\label{4.19a}\\
&&2\overline{\chi}_{ik}=
\partial_s\langle\widetilde{\xi}_i\widetilde{\xi}_k\rangle\label{4.20bb}
\end{eqnarray}
In fact, one can see that the equations \eqref{4.15a}-\eqref{4.20bb} are very similar to that of DL(2) (\ref{4.15})-(\ref{4.17});
the difference is: the same coefficients as in DL(2) appear in the previous approximations of DL(3).

For DL(4) we impose two restrictions  $\overline{\vV}_0\equiv 0$ and $\overline{\vV}_1\equiv 0$ and derive (by a similar procedure) the equation
\begin{eqnarray}
&&\left(\partial_s+ \overline{\vV}_2\cdot\nabla\right)\overline{a}_{0}=
\frac{\partial}{\partial x_i}\left(
\overline{\chi}_{ik}\frac{\partial\overline{a}_0}{\partial x_k}\right),
\label{4.17c}
\end{eqnarray}
with the same $\overline{\vV}_2$ and $\overline{\chi}_{ik}$ as in DL(3).

In general, the comparison between the averaged equations for DL(1)-(4) shows that the same coefficients in DL($n$) appear in the equations of the next order (in $\varepsilon$) in DL($n-1$).
The higher approximations DL(5), \emph{etc.} can be derived similarly, however the calculations become too cumbersome.

\section{One-parametric families of solutions}

One can see, that, in all presented above distinguished limit equations, the meaning of key \emph{parameter $\varepsilon$ is not uniquely  defined in terms of $\omega$ or $St$}.
In  DL(2)  $\varepsilon_1=\varepsilon$, $\varepsilon_2=\varepsilon^2$  in (\ref{exact-6a}),
which imposes the link $\alpha$, $\beta$ and $\varepsilon$:
\begin{equation}
\alpha=1-2\beta,\quad \beta=(1-\alpha)/2,\quad \varepsilon=1/\omega^{(\alpha+1)/2}
\label{main-eq111}
\end{equation}
Hence, each distinguished limit solution of (\ref{exact-6a}), obtained in terms of $\varepsilon$, produces an infinite number of \emph{parametric solutions}, one solution for any real number $\beta<1$ (or for any real number $\alpha>-1$).
Those solutions are mathematically similar but physically different, since they correspond to different magnitudes of given velocities $\vv^*$ \eqref{variables} and different slow-time variables $s=t/\omega^\alpha$. There are two ways of choosing $\alpha$ and $\beta$.

Given $\alpha$: When the slow time-scale $s$ is given in the prescribed function $\vu(\vx, s,\tau)$ \eqref{variables1}, then it defines $\alpha$; then $\beta$ follows from \eqref{main-eq111}.

Given $\beta$: Alternatively, when the velocity amplitude of $U\omega^\beta$ is given, then it defines $\beta$ and we have to calculate $\alpha$ from \eqref{main-eq111}.

Some interesting sets of $\alpha$ and $\beta$ are:

(i) If $\widetilde{\vu}$ is given as the function of variables $\tau=\omega t$ and $s=t$, then  $\alpha=0$, $\beta=1/2$; hence the velocity $\vv^*\sim O(\sqrt{\omega})$ (\ref{variables}), and in \eqref{basic-4aa} $\varepsilon=1/\sqrt\omega$.

(ii) The most frequently considered case is $\vv^*/U\sim O(1)$, then $\alpha=1$, $\beta=0$, $s=t/\omega$, and $\varepsilon=1/\omega$.

 (iii) Rather exotic possibility corresponds to the case $\alpha=\beta =1/3$. Here, $s=t/\sqrt[3]\varepsilon$, $\varepsilon=\omega^{-2/3}$ and velocity $\vv^*/U=O(\sqrt[3]\omega)$. Such a scaling is required, for example,  if a particular slow time-scale $s=t/\sqrt[3]\omega$ is prescribed in $\widetilde{\vu}$.

 (iv) The above results for DL(2) are particularly striking for the case of  $\vu=\widetilde{\vu}(\vx,\tau)$, when the given velocity does not contain any slow time-scale $s$ and represents purely periodic velocity oscillations. An intricate question arises: how to choose any particular scale of $s$?
 Our answer is: in this case the slow time-scale is uniquely determined by the magnitude $O(\omega^\beta)$ of $\vv^*/U$. For every given value of $\beta$ we obtain the related slow time variable $s=t/\omega^{(1-2\beta)}$. It should be accepted that solutions with different $\beta$ are physically different since they correspond to different orders of a prescribed velocity field and different slow time-scales $s$.

 (v) The functional dependence \eqref{main-eq111} between $\alpha$ and $\beta$ is physically natural: the increasing of the amplitude of a given velocity (increasing $\beta$) leads to the decreasing of the slow time-scale (decreasing $\alpha$).

In general, the transformations similar to (\ref{main-eq111}) produce an infinite number of \emph{parametric solutions} for each case DL(1-4).

\section{Does pseudo-diffusion represent a mystery?}

The inevitable appearance of the \emph{pseudo-diffusion} terms in our averaged equations does represent an intriguing result.
We have introduced the term  {\emph{pseudo-diffusion }} on the grounds of the following arguments:

(i) the evolution of $\overline{a}$ is described by an advection-pseudo-diffusion-type equation (\ref{4.21}),
where the matrix of pseudo-diffusion coefficients $\overline{\chi}_{ik}$ can be positive, negative or sign-indefinite;
it also can change its sign in space and time;\

(ii) the equation (\ref{4.21}) is valid only for the regular
asymptotic expansions (\ref{4.22})-(\ref{4.22aa}), while the classical diffusion is often affiliated with the singular expansions;

(iii) for the most interesting case DL(2) the pseudo-diffusion  term represents a known `source' term in the second approximation \eqref{4.17}; for DL(4) the pseudo-diffusion does appear already in the leading approximation \eqref{4.17c}, although its evolution is described by `very' slow time $s=t/\omega^3$;

(iv) The possibility of the secular growth (in $s$) of averaged solutions due to pseudo-diffusion represents a central open question. Such secular terms could appear for the pseudo-diffusion matrix $\overline{\chi}_{ik}$ \eqref{4.20}, which is constant or monotonically increasing in $s$, but they are unlikely to appear if $\overline{\chi}_{ik}$ is oscillating in $s$.
If such a growth does appear, then the sufficiency of two time-scales only should be reconsidered.

(v) the appearance of pseudo-diffusion matrix as a \emph{Lie-derivative of an averaged quadratic displacement tensor} \eqref{4.20}, \eqref{4.20b} represents the  result of straightforward calculations,  which still requires qualitative understanding. We propose the  following example as a first step of such understanding.

\emph{\underline{Example of rigid-body oscillations}:} The simple and  instructive example of a prescribed velocity is a rigid-body oscillations of a medium, where $\vu=\widetilde{\vu}(s,\tau)$. Due to the vanishing of all spatial derivatives, all approximations of a drift velocity also vanish $\overline{\vV}_k\equiv 0$ ($k=0,1,2,\dots$) and the DL(4) equation  \eqref{4.17c} takes a simplified form
\begin{eqnarray}
&&\partial_s\overline{a}_{0}=
\frac{\partial}{\partial x_i}\left(
\overline{\chi}_{ik}\frac{\partial\overline{a}_0}{\partial x_k}\right),\quad \overline{\chi}_{ik}=\partial_s\langle\widetilde{u}_i\widetilde{u}_k\rangle
\label{4.17d}
\end{eqnarray}
There are few observations for this equation:

(a) When $\widetilde{\vu}=\widetilde{\vu}(\tau)$  (purely periodic oscillations of a velocity), then $\overline{\chi}_{ik}\equiv 0$, which leads to a steady distribution $\overline{a}_{0}(\vx)$;
hence, any purely periodic rigid-body oscillations do not affect the averaged field $\overline{a}_{0}$.

(b) In order to obtain $\overline{\chi}_{ik}=C_{ik}$ (with $C_{ik}$ a constant tensor), one has to take $\widetilde{\vu}=\sqrt{s}\widetilde{\vf}(\tau)$, where $\widetilde{\vf}(\tau)$ is an arbitrary tilde-function; we note that for  positive matrix $C_{ik}$ we have a standard diffusion equation, for  negative $C_{ik}$ -- an `anti-diffusion' equation, and for a sign-indefinite matrix -- a pseudo-diffusion equation.

(c) One can see, that for  positive matrix $C_{ik}$ the related evolution of $\overline{a}_{0}$ is described by  standard diffusion equation \eqref{4.17d}. However, in a contrast to classical diffusion, the mechanism of this evolution is different: this evolution is due to the growth of a spatial oscillatory amplitude as $\sqrt{s}$;
hence, the described phenomenon (which looks like diffusion) actually represents an advection (by the increasing in amplitude oscillations), which is `superimposed' with the average operation. Hence, we have arrived to the following striking result: the superposition of advection due to rigid-body oscillations and the eulerian averaging operation  can lead to the evolution of an averaged solution, that is mathematically equivalent to the classical diffusion.

(d) For $\widetilde{\vu}(s,\tau)$ being a periodic function of $s$ (which represents modulated oscillations), the sign of $\overline{\chi}_{ik}$ is periodically changing, mathematical properties of such  PDE are unknown, however one might expect that such an equation has  bounded solutions (secular growth is absent). The case when the eigenvalues of $\overline{\chi}_{ik}$ have different signs, has not been systematically studied yet, see \cite{Mizrahi}.

(e) For understanding  of applicability of \eqref{4.17d} one  should keep in mind that all our dimensionless functions and their derivatives are of order one, hence, the consideration of solutions with large gradients (\emph{e.g.} a point source solution) represents an invalid operation.

(f) We emphasise, that for non-zero drift velocities all the  equations \eqref{4.17}, \eqref{4.17a} and \eqref{4.17c} contain additional spatial derivatives (in comparison with \eqref{4.17d}). In particular,  pseudo-diffusion matrix $\overline{\chi}_{ik}$ contains spatial derivatives (as parts of the Lie-derivative); this intriguing appearance of pseudo-diffusion remains completely unexplained from a physical viewpoint.

\section{Examples}

All the above averaged equations have been derived for arbitrary function $\widetilde{\vu}(\vx, s,\tau)$, therefore, the number of interesting examples can be significant. In this section we consider only five instructive classes of $\widetilde{\vu}(\vx, s,\tau)$.
\vskip 2mm

\textbf{\emph{Example 1.} \emph{The superposition of two modulated oscillatory fields:}}
\begin{eqnarray}
&&\widetilde{\vu}(\vx,s,\tau)=\overline{\vp}(\vx,s)\sin\tau+
\overline{\vq}(\vx,s)\cos\tau  \label{Example-1-1}
\end{eqnarray}
where
$\overline{\vp}$ and $\overline{\vq}$ are arbitrary bar-functions. The straightforward calculations lead to
$
[\widetilde{\vu},\widetilde{\vxi}]=[\overline{\vp},\overline{\vq}].
$
Equations (\ref{4.18}), (\ref{4.19}), \eqref{4.20} yield
\begin{eqnarray}
&&\overline{\vV}_0=\frac{1}{2}[\overline{\vp},\overline{\vq}],\quad
\overline{\vV}_1\equiv 0,
\label{Example-1-4}\\
&&\overline{\vV}_2=\frac{1}{8}\left([\overline{\vP},\overline{\vp}]+[\overline{\vQ},\overline{\vq}]\right)
-\frac{1}{4}\left([\overline{\vp}_s,\overline{\vp}]+[\overline{\vq}_s,\overline{\vq}]\right)+\label{Example-1-4a}\\
&&+\frac{1}{4}\left(\overline{\vp}\,\Div\overline{\vP}'+\overline{\vq}\Div\overline{\vQ}'
+\overline{\vP}'\Div\overline{\vp}+\overline{\vQ}'\Div\overline{\vq}\right),\nonumber\\
&&\overline{\vP}\equiv[\overline{\vV}_0,\overline{\vp}],\quad
\overline{\vQ}\equiv[\overline{\vV}_0,\overline{\vq}],\quad
\overline{\vP}'\equiv\overline{\vp}_s-\overline{\vP},\quad
\overline{\vQ}'\equiv\overline{\vq}_s-\overline{\vQ},\nonumber\\
&&\langle\widetilde{\xi}_i\widetilde{\xi}_k\rangle=
\frac{1}{2}(\overline{p}_i \overline{p}_k + \overline{q}_i \overline{q}_k)\label{Example-1-5}
\end{eqnarray}
The expression for the \emph{pseudo-diffusion} matrix $\overline{\chi}_{ik}$ follows after the substitution of (\ref{Example-1-5}) into
(\ref{4.20}). The expression $\widetilde{\vu}$ (\ref{Example-1-1}) is general enough to produce any given function
$\overline{\vV}_0(\vx,t)$. Indeed, in order to obtain $\overline{\vp}(\vx,t)$ and $\overline{\vq}(\vx,t)$ one has to solve the equation
\begin{eqnarray}\label{bi-linear}
[\overline{\vp},\overline{\vq}]=(\overline{\vq}\cdot\nabla)\overline{\vp}-
(\overline{\vp}\cdot\nabla)\overline{\vq}=2\overline{\vV}_0(\vx,s)
\end{eqnarray}
which represents an undetermined bi-linear PDE-problem for two unknown functions $\overline{\vp}(\vx,t)$ and $\overline{\vq}(\vx,t)$.
\vskip 2mm

\textbf{\emph{Example 2.} \emph{ Stokes drift:}}

The dimensionless plane velocity field for an inviscid incompressible fluid is (see \cite{Stokes, Lamb, Debnath})
\begin{eqnarray}
\widetilde{\vu}=Ae^{ky}\left(\begin{array} {c} \cos(kx-\tau)\\ \sin (kx-\tau)\end{array}\right),\quad (x,y)\equiv(x_1,x_2)
\label{Example-3-1}
\end{eqnarray}
where for our consideration one can take $A=A(s)$ and $k=k(s)$, however we choose $A$ and $k$ as constants; however, for our consideration they can be chosen as arbitrary functions of $s$. The fields $\overline{\vp}(x,y)$, $\overline{\vq}(x,y)$ (\ref{Example-1-1}) are
\begin{eqnarray}
\overline{\vp}=Ae^{ky}\left(\begin{array}{c} \sin kx\\ -\cos kx\end{array}\right),\quad
\overline{\vq}=Ae^{ky}\left(\begin{array}{c} \cos kx\\ \sin kx\end{array}\right)
\label{Example-3-2}
\end{eqnarray}
The calculations of (\ref{Example-1-4}) yield
\begin{eqnarray}
\overline{{\vV}}_0=k A^2 e^{2ky}\left(\begin{array}{c} 1\\ 0\end{array}\right), \quad  \overline{{\vV}}_1\equiv 0
\label{Example-3-3}
\end{eqnarray}
which represent the classical Stokes drift and the first correction
to it (which vanishes).  For brevity, the explicit formula for $\overline{{\vV}}_2$  is not given here.
Further calculations show that
\begin{eqnarray}\nonumber
&&\overline{\chi}_{ik}=-\overline{\chi}
\begin{pmatrix}
0 & 1\\
1 & 0
\end{pmatrix},\quad\text{with}\quad \overline{\chi}\equiv\frac{1}{4}k^2 A^4 e^{3ky}
\end{eqnarray}
One can see that  eigenvalues $\overline{\chi}_1=-\overline{\chi}$ and $\overline{\chi}_2=\overline{\chi}$
correspond to strongly anisotropic pseudo-diffusion.
The averaged equation (\ref{4.21}) (with an error $O(\varepsilon^3)$) can be written as
\begin{eqnarray}
&&\overline{a}_{s}+(\overline{V}_0+\varepsilon^2 \overline{V}_2) \overline{a}_x=
    \varepsilon^2(\overline{\chi}_y \overline{a}_{x}+\overline{\chi}\, \overline{a}_{xy})\label{Example-3-5}\\
    &&\overline{a}=\overline{a}_0+\varepsilon\overline{a}_1+\varepsilon^2\overline{a}_2\nonumber
\end{eqnarray}
where $\overline{V}_0$ and $\overline{V}_2$ are $x$-components of corresponding velocities (their $y$-components
vanish). This equation has an exact solution $\overline{a}=\overline{a}(y)$ where $\overline{a}(y)$ is an arbitrary function, which is not affected  by pseudo-diffusion.
\vskip 2mm

\textbf{\emph{Example 3.}  \emph{A spherical `acoustic' wave:}}

The main term for velocity potential for an outgoing spherical acoustic wave is
\begin{eqnarray}
&& \widetilde{\phi}=\frac{A}{r}\sin(kr-\tau)\label{Example-4-1}
\end{eqnarray}
where $A=\const$, $k=\const$, and $r$ are an amplitude, a wavenumber, and a radius in a spherical coordinate system. The velocity is
purely radial and has a form (\ref{Example-1-1})
\begin{eqnarray}
&& \widetilde{u}=\overline{p}\sin\tau+\overline{q}\cos\tau,\\
&&\overline{p}=A\left(\frac{1}{r^2}\cos kr+\frac{k}{r}\sin kr\right),\quad\overline{q}=A\left(-\frac{1}{r^2}\sin
kr+\frac{k}{r}\cos kr\right)\label{Example-4-1a}
\end{eqnarray}
where $\widetilde{u}, \overline{p}$, and $\overline{q}$ are radial components of corresponding vector-fields. One can find that the
fields $\widetilde{\vxi}$ and $[\widetilde{\vu},\widetilde{\vxi}]$ are also purely radial; the radial component for the
commutator is
\begin{eqnarray}
&& \widetilde{\xi} \widetilde{u}_r-\widetilde{u}\widetilde{\xi}_r=A^2 k^3/r^2\label{Example-4-3}
\end{eqnarray}
where $\widetilde{\xi}$ is the radial component of $\widetilde{\vxi}$ and subscript $r$ stands for the radial derivative. The drift
(\ref{Example-1-4}) is purely radial with
\begin{eqnarray}
&&\overline{V}_0=\frac{A^2 k^3}{2\,r^2},\quad
\overline{V}_1=0,\quad\overline{V}_2=\frac{A^4k^5}{16r^4}\left(3k^2-\frac{5}{r^2}\right)
\label{Example-4-5}
\end{eqnarray}
It is interesting, that $\overline{V}_0$ formally coincides with the velocity, caused by a point source in an
incompressible fluid, and (for small $r$) the value of  $\overline{V}_2$
dominates over $\overline{V}_0$, so the series is likely to be diverging at $r=0$. Further calculations yield
\begin{eqnarray}
&&\langle\xi^2\rangle=\frac{A^2}{2r^2}(k^2+1/r^2),\quad \overline{\chi}=A^4k^5/4r^2>0
\label{Example-4-6}
\end{eqnarray}
where $\overline{\chi}$ stands for the only nonzero $rr$-component of $\overline{\chi}_{ik}$. One can see, that in this
case pseudo-diffusion appears as ordinary diffusion.

\vskip 2mm

\textbf{\emph{Example 4.}  \emph{ $\overline{\vV}_1$-drift:}}

It is interesting to consider such flows for which the classical expression for a drift vanishes: $\overline{\vV}_0\equiv 0$, but $\overline{\vV}_1\neq 0$. Let the velocity field
be a superposition of two standing waves of frequencies $\omega$ and $2\omega$:
\begin{eqnarray}
&&\widetilde{\vu}(\vx,t,\tau)=\overline{\vp}(\vx,t)\sin\tau+\overline{\vq}(\vx,t)\sin 2\tau  \label{Example-5-1}\\
&&[\widetilde{\vu},\widetilde{\vxi}]=\frac{1}{2}[\overline{\vp},\overline{\vq}](2\cos\tau\sin 2\tau-\cos
2\tau\sin\tau)\label{Example-5-3}
\end{eqnarray}
Hence (\ref{4.18}) yields
\begin{eqnarray}\label{Example-5-4}
\overline{\vV}_0=\frac{1}{2}\langle[\widetilde{\vu},\widetilde{\vxi}]\rangle\equiv 0,\quad
\overline{\vV}_1=\frac{1}{3}\langle[[\widetilde{\vu},\widetilde{\vxi}],\widetilde{\vxi}]\rangle=
\frac{1}{8}[[\overline{\vp},\overline{\vq}],\overline{\vp}]
\end{eqnarray}
These expressions produce infinitely many examples of the flows with $\overline{\vV}_1$-drift. At the same time \eqref{Example-5-4} shows that for a single standing wave the $\overline{\vV}_1$-drift is absent.

\textbf{\emph{Example 5.}  \emph{ Chaotic averaged `lagrangian' dynamics for $\overline{\vV}_0$:}}

Let the solenoidal/incompressible velocity (\ref{Example-1-1})
be
\begin{eqnarray}
\overline{\vp}=\left(\begin{array}{c}
  \cos y \\
  0 \\
  \sin y
\end{array}\right), \qquad
\overline{\vq}=\left(\begin{array}{c}
  a\sin z \\
  b\sin x+a\cos z \\
  b\cos x
\end{array}\right)
\end{eqnarray}
where  $(x,y,z)$ are  cartesian coordinates; $a, b$ are constants. Either of these fields, taken separately, produces
simple integrable dynamics of particles. The calculations yield
\begin{eqnarray}
\overline{\vV}_0=\left(\begin{array}{c}
  -a\sin y\sin x-2b\sin y\cos z \\
  b\sin z\sin y-a\cos x\cos y \\
  b\cos z\cos y+2a\sin x\cos y
\end{array}\right),
\end{eqnarray}
The computations of the averaged `lagrangian' dynamics for $d\overline{\vx}_0/ds=\overline{\vV}_0$ (based on (\ref{exact-1a}) for DL(2)) were performed by Prof. A.B.Morgulis (private communications). He has shown that this steady averaged flow exhibits chaotic dynamics of particles.
In particular, positive
Lyapunov exponents have been observed. Hence, the drift created by a simple oscillatory field can
produce complex \emph{averaged `lagrangian' dynamics}.

\section{Discussion}


\emph{Struggle for natural notations:}
The introducing of convenient and logical notations represents an important part of this paper. This topic is always underestimated, however natural notations are especially helpful in the presence of cumbersome calculations.

\vskip 1mm
\emph{Averaged and oscillatory parts of solutions:}
The analysis, presented in the paper,  produces not only the averaged parts of solutions but also full solutions.
Indeed, in Appendix one can find the oscillatory parts of solutions in all considered approximations, see \eqref{A01-appr-1}, \eqref{01-appr-7}, \eqref{2-appr-2}, \eqref{2-appr-11}, \eqref{4-appr-5}.

\vskip 1mm
\emph{Does three-timing method really needed?}
 This paper is aimed to create a general viewpoint on distinguished limits, drifts, and pseudo-diffusion based on the rigorous implementation of the two-timing method. We do not use any additional suggestions and assumptions, hence any contradictions (either mathematical or physical) in further results (which follow from our averaged equations and solutions) can be explained by insufficiency of two time-scales only. Indeed, the presence of scaling parameters, such as $St$, $\varepsilon_1$, and $\varepsilon_2$, allows one to introduce an infinite number of additional time-scales.
From this perspective the considered problem can be seen as a test for sufficiency of the two-timing method.
In particular, due to the presence of pseudo-diffusion, some secular (in $s$) terms could appear in solutions of \eqref{4.17}, \eqref{4.17a} and \eqref{4.17c}.
If it is proven, then one may suggest that the two-timing method fails at the orders of approximations, where  such a secular growth appears.
In this case further time-scales (additional to $\tau$ and $s$) can be introduced, which requires the systematic development of three-timing \emph{etc.} methods. However, any  mathematically systematic method, which allows to derive the averaged equations with three or more time-scales from an original PDE, is still unknown.

\vskip 1mm
\emph{Mathematical justification of the two-timing method:}
One can rewrite the approximate solution \eqref{4.22aa} (along with its tilde-parts given in Appendix) back to the original time variable $t$ and substitute it into the exact original equation (\ref{exact-1}): then a small residual (a nonzero right-hand-side in (\ref{exact-1})) appears.
The two-timing method (if it used formally) allows to produce an approximate solution with a residual
as small as required by an user. However, the next logical step is more challenging: one has to prove that the solution of the equation with a small residual (instead of zero in the right-hand side) is close to the exact solution.
For the two-timing procedure such proofs had been performed for the leading approximation of solution by \cite{Simonenko, Levenshtam} in the problem of vibrational convection.
Similar justifications
for other equations with oscillatory coefficients are not known yet.

\vskip 1mm
\emph{The interplay between a velocity amplitude and slow time-scale:}
A different (from the presented in this paper) approach is given in \cite{Yudovich, VladimirovDr1} where \emph{an inspection procedure} for calculation of distinguished paths is proposed and the concept of `strong',  `moderate', and `weak' oscillations (or vibrations) is introduced.
The advantage of the approach presented in this paper is the additional possibility to vary the slow time-scale $s$. It makes the structure of asymptotic solutions more flexible and  allows to consider  broader classes of asymptotic solutions. However, from a `physical viewpoint' some of our results can appear as `unexpected' and `paradoxical'.
Say, in the case $\vv^*/U\sim O(1)$ in \eqref{variables1}, which looks `physically natural',  one should take $\alpha=1$, $\beta=0$, $s=t/\omega$, and $\varepsilon=1/\omega$ \eqref{main-eq111}. Then the appearance of the variable $s=t/\omega$ may not be seen as `natural'.
From another side, if one takes $s=t$ (which also looks  `physically natural'), then it must be $\vv^*/U\sim O(\sqrt{\omega})$, that, again, could be viewed as physically `artificial'. Such an interplay between the scale of velocity and slow time-scale does create some ample confusions and misunderstandings.

\vskip 1mm
\emph{The most striking case -- purely periodic oscillations:}
The results for DL(2) \eqref{main-eq111}  are particularly striking for the case of  purely periodic velocity oscillations, when $\vu=\widetilde{\vu}(\vx,\tau)$ is independent of $s$. Here the most intricate question is: how to choose any particular scale of $s$?
We have an answer: the slow time-scale is uniquely determined by the magnitude $O(\omega^\beta)$ of $\vv^*$. For every value of $\beta$ we obtain the related DL(2)  slow time $s=t/\omega^{(1-2\beta)}$. It should be accepted that solutions with different $\beta$ are physically different, since they correspond to different orders of the prescribed velocity field $\vv$.

An additional `naive' question, which could be asked here, is: why do we need to introduce any slow time-scale $s$ at all? The answer to this question is: if for the coefficient $\vu=\widetilde{\vu}(\vx,\tau)$ we consider only a solution of similar structure $a=a(\vx,\tau)$, then we can overlook a broad class of asymptotic solutions.
A well-known and clarifying example here is the solutions of Mathieu equation, which represents an ODE with purely periodic coefficient, see \cite{mathieu}. The slow time-scale, which appears there (say, in the slowly growing solutions, related to the parametric resonance), is proportional to the amplitude of frequency modulation. Our case looks somehow similar: the slow time-scale appears to be related to the amplitude of velocity oscillations. We believe that such a relation between the amplitude of purely oscillating coefficients and the slow time-scale represents a general property of related PDEs and ODEs.

\vskip 1mm
\emph{On the term `pseudo-diffusion':}
 This term has been in a diverse use in various applied disciplines, some of them remote from fluid dynamics and mathematical physics, say in Chemistry, Geology, \emph{etc}. For example, its definition from geology is:
``Mixing of thin superposed layers of slowly accumulated marine sediments by the action of water motion or subsurface organisms''
(see McGraw-Hill Dictionary of Scientific and Technical Terms, 6E,  2003 by The McGraw-Hill Companies, Inc.).
However, there is one purely mathematical definition, where `pseudo-diffusion' appears  as the `hyperbolic type' spatial operator $u_t=u_{xx}-f(t)u_{yy}$ with $f(t)>0$, see \cite{Mizrahi}. This definition is qualitatively close to one case (Stokes drift) which appears in this paper. At the same time, we have extended the meaning of `pseudo-diffusion' by including all possible cases such as $u_t=u_{xx}+f(x,y,t)u_{yy}$ with the function $f(x,y,t)$ changing its sign in space and time.

\vskip 1mm
\emph{Averaged `lagrangian' dynamics vs. exact lagrangian dynamics?}
\emph{Example 5} demonstrates that a drift can produce chaotic averaged dynamics which in this case cannot be directly affiliated with material particles..
This result brings on a number of questions, such as:
(i) what is the relationship between the lagrangian chaotic motions for the original dynamical system and `lagrangian' chaotic motions for the averaged one?
(ii) how can  chaotic averaged `lagrangian' motion and  pseudo-diffusion complement each other?
(iii) how can `averaged chaos', induced by a drift,  be used in the theory of mixing?
(iv) let the averaged `lagrangian' dynamics be chaotic, then how can the related results by
\cite{Arnold,Aref,Ottino,Wiggins,Chierchia}, and many others be used in applications.

Another related example could be the ABC-flow by \cite{Frish}, however, to make it relevant to the topic of this paper we need to find such an oscillating flow which has the ABC-flow as its drift.

\vskip 1mm
\emph{Additional reading materials:}
For an interested reader: all the presented in this paper calculations as well as the calculations for different distinguished limits and a number of additional (to Sect.7) examples are given in the arXiv papers by \cite{VladimirovDr1, VladimirovDr2}, which are quoted below as I and II.

\vskip 1mm
\emph{Is our set of distinguished limits DL(n) complete?}
 The key question can be asked: does the described in this paper set of distinguished limits and parametric solution represent a full set of asymptotic solutions of  \eqref{exact-6a}, \eqref{paths} obtainable by the two-timing method? The answer is unknown, but most likely some other solutions do exist; a few attempts to identify additional solutions are given in I and II.

\vskip 1mm
\emph{Simplification of $\tau$-periodicity:}
All results of this paper have been obtained for the class of $\tau$-periodic functions, which is self-consistent. One can consider more general
classes of quasi-periodic, non-periodic, or chaotic solutions. The discussion on this topic is given in I, II.
However, it is worth to understand the properties of $\tau$-periodic oscillations, in order not to relate these properties exclusively to more general solutions.
At the same time, the relative simplicity of calculations for the  $\tau$-periodic solutions allows to obtain some advanced results which can serve as a guidance for making assumptions on the properties of more general solutions.

\vskip 1mm
\emph{Lagrangian drift vs. Eulerian drift.}
It is worth to calculate the classical (lagrangian) drift directly by solving (\ref{exact-1a}) with the use of the same two-timing method. The drift, obtained in this paper, appeared as the result of eulerian average operation; hence, its comparison with the classical drift represents an open problem. This topic requires a separate study which has been started in I, II.

\vskip 1mm
\emph{Advection of vectorial admixture.}
The averaged equations for a passive vectorial admixture are presented in I, II.
These equations are closely linked to the problem of \emph{kinematic $MHD$-dynamo}, see \cite{Moffatt}.
It is physically apparent, that for the majority of shear drift velocities $\overline{\vV}(\vx,s)$ the averaged stretching of `material
lines'  produces the linear in $s$ growth of a magnetic field $|\overline{\vh}|\sim s$. At the same time, for the averaged flows with
exponential stretching of averaged `material lines' these examples will inevitably show the exponential growth.

Another closely related research topic is the `advection' of an active vectorial admixture (vorticity). The vortex dynamics of oscillating flows has been studied in \cite{VladProc}. An interesting phenomenon here is the  \emph{Langmuir circulations}, see
\cite{Craik0}, which has been recently analyzed from a new perspective by \cite{VladimirovL}.
All these topics are worth to be studied by the approach presented in this paper.

\begin{acknowledgments}
The author is
grateful to Prof. K.I. Ilin for the checking of calculations and to Prof. A.B. Morgulis for the computing Example 5. The author wants to express special thanks to Profs. A.D.D. Craik and H.K. Moffatt for reading this manuscript and making useful critical remarks, also many thanks to Profs. M.E. McIntyre, T.J. Pedley, M.R.E. Proctor, and D.W. Hughes  for helpful discussions. This research is supported by the grant IG/SCI/DOMS/16/13 from Sultan Qaboos University, Oman.

\end{acknowledgments}

\appendix

\section{DL(2): Asymptotic procedure and detailed solution.  \label{sect04}}

First, we list the properties of $\tau$-differentiation and tilde-integration \eqref{ti-integr} which are intensely used in the calculations below.
For $\tau$-derivatives it is clear that
\begin{eqnarray}
f_\tau=\overline{{f}}_\tau+\widetilde{f}_\tau=\widetilde{f}_\tau, \quad
\langle f_\tau\rangle=\langle\widetilde{f}_\tau\rangle=0 \label{oper-6}
\end{eqnarray}
The product of two tilde-functions $\widetilde{f}$ and $\widetilde{g}$ forms a general $\tau$-periodic function:
$\widetilde{f}\widetilde{g}\equiv F$, say. Separating the tilde-part $\widetilde{F}$ from $F$ we
write
\begin{eqnarray}
&&\widetilde{F}=F-\langle F\rangle
=\widetilde{f}\widetilde{g}-\langle\widetilde{f}\widetilde{g}\rangle=
\widetilde{\widetilde{f}\widetilde{g}}\equiv\{\widetilde{f}\widetilde{g}\},\quad \widetilde{F}\equiv {F}
\label{oper-5}
\end{eqnarray}
where the introducing of braces for the tilde parts of a function is aimed to avoid the cumbersome two-level tilde notation.
As the average operation represents the integration over $\tau$, then for products
containing tilde-functions $\widetilde{f},\widetilde{g},\widetilde{h}$ and their derivatives we have
\begin{eqnarray}
&&\langle\widetilde{f}\widetilde{g}_\tau\rangle=\langle(\widetilde{f}\widetilde{g})_\tau\rangle-
\langle\widetilde{f}_\tau\widetilde{g}\rangle=-\langle\widetilde{f}_\tau \widetilde{g}\rangle=-
\langle\widetilde{f}_\tau g\rangle
\label{oper-9}\\
&&\langle\widetilde{f}_\tau\widetilde{g}\widetilde{h}\rangle+\langle\widetilde{f}\widetilde{g}_\tau\widetilde{h}\rangle+
\langle\widetilde{f}\widetilde{g}\widetilde{h}_\tau\rangle=0
\label{oper-9a}\\
&&\langle\widetilde{f}\widetilde{g}^\tau\rangle=\langle(\widetilde{f}^\tau\widetilde{g}^\tau)_\tau\rangle-
\langle\widetilde{f}^\tau\widetilde{g}\rangle=-\langle\widetilde{f}^\tau \widetilde{g}\rangle=-
\langle\widetilde{f}^\tau g\rangle
\label{oper-10}
\end{eqnarray}
which can be seen as different versions of integration by parts. Similarly, for the commutators we have
\begin{eqnarray}
&&\langle[\widetilde{\va},\widetilde{\vb}_\tau]\rangle=-\langle[\widetilde{\va}_\tau,\widetilde{\vb}]\rangle=-
\langle[\widetilde{\va}_\tau, \vb]\rangle,\
\langle[\widetilde{\va},\widetilde{\vb}^\tau]\rangle=-\langle[\widetilde{\va}^\tau,\widetilde{\vb}]\rangle=-
\langle[\widetilde{\va}^\tau, \vb]\rangle
\label{oper-15}
\end{eqnarray}
Now, we describe the obtaining of the solution of equation \eqref{exact-6c} for $n=2$
\begin{eqnarray}
&& a_\tau+\varepsilon^2 a_s+\varepsilon(\widetilde{\vu}\cdot\nabla)\,a=0
\label{Abasic-2}
\end{eqnarray}
The substitution of (\ref{basic-4aa}) into this equation produces the equations of successive approximations
\begin{eqnarray}
&&a_{0\tau}=0 \label{Abasic-5}\\
&&a_{1\tau}=-(\widetilde{\vu}\cdot\nabla)\, a_0\label{Abasic-6}\\
&&a_{n\tau}=-(\widetilde{\vu}\cdot\nabla)\,
a_{n-1}-\partial_s\, a_{n-2},\quad\partial_s\equiv\partial/\partial s,\quad n=2,3,4,\dots
\label{Abasic-7}
\end{eqnarray}
The separation of the tilde-parts $\widetilde{a}_k$, for $k=0,1,2,3,4$ produces the explicit recurrent expressions
\begin{eqnarray}
&&\widetilde{a}_0\equiv 0,\label{4.11a}\\
&& \widetilde{a}_1= -(\widetilde{\vxi}\cdot\nabla)\,\overline{a}_0,\label{A4.11}\\
&&\widetilde{a}_{2}=-(\widetilde{\vxi}\cdot\nabla)\, \overline{a}_1
-\{(\widetilde{\vu}\cdot\nabla)\,\widetilde{a}_1\}^\tau, \label{A4.12}\\
&&\widetilde{a}_{3}=-(\widetilde{\vxi}\cdot\nabla)\,\overline{a}_2
-\{(\widetilde{\vu}\cdot\nabla)\,\widetilde{a}_2\}^\tau-\widetilde{a}_{1s}^\tau, \label{A4.13}\\
&&\widetilde{a}_{4}=-(\widetilde{\vxi}\cdot\nabla)\,\overline{a}_3-
\{(\widetilde{\vu}\cdot\nabla)\,\widetilde{a}_3\}^\tau-\widetilde{a}_{2s}^\tau, \label{A4.14}
\end{eqnarray}
Further calculations and transformation with the use of \eqref{oper-6}-\eqref{oper-15} show that the bar-parts $\overline{a}_k$ satisfy the equations \eqref{4.15}-\eqref{4.20b}.
Let us present the derivations of \eqref{4.11a}-\eqref{A4.14} and \eqref{4.15}-\eqref{4.20b}.

\underline{\emph{The zero-order equation}}  (\ref{Abasic-5})  is:
\begin{eqnarray}
&&a_{0\tau}=0\label{A01-appr-1}
\end{eqnarray}
The substitution of $a_{0}=\overline{a}_0(\vx,s)+\widetilde{a}_0(\vx,s,\tau)$ into (\ref{A01-appr-1}) gives
$\widetilde{a}_{0\tau}= 0$. Its tilde-integration \eqref{ti-integr}   produces the unique (inside the
tilde-class) solution $\widetilde{a}_0\equiv 0$. At the same time, (\ref{A01-appr-1}) does not impose any
restrictions on $\overline{a}_0(\vx,s)$, which must be determined from the next approximations. Thus the results
derivable from (\ref{A01-appr-1}) are:
\begin{eqnarray} &&\widetilde{a}_0(\vx,s,\tau)\equiv 0;\quad \overline{a}_0(\vx,s) \ \text{is not defined}\label{A01-appr-1a}
\end{eqnarray}

\underline{\emph{The first-order equation} } (\ref{Abasic-6})  is
\begin{eqnarray}
&&a_{1\tau}=-(\widetilde{\vu}\cdot\nabla)\, a_0 \label{A01-appr-3}
\end{eqnarray}
The use of $\widetilde{a}_0\equiv 0$ and $\overline{a}_{1\tau}\equiv 0$ reduces (\ref{A01-appr-3}) to
the equation $\widetilde{a}_{1\tau}=-(\widetilde{\vu}\cdot\nabla)\,\overline{a}_0$. Its tilde-integration gives the unique solution
\begin{eqnarray}
&&\widetilde{a}_{1}=-(\widetilde{\vxi}\cdot\nabla)\,\overline{a}_0\quad \text{where}\quad \vxi\equiv\vu^\tau\label{01-appr-7}
\end{eqnarray}
 Hence,
\begin{eqnarray}
&&a_1=\overline{a}_1 -(\widetilde{\vxi}\cdot\nabla)\,\overline{a}_0
\label{01-appr-8}
\end{eqnarray}
where $\overline{a}_0(\vx,s), \overline{a}_1(\vx,s)$ are not defined.

\underline{\emph{The second-order equation}} ((\ref{Abasic-7}) for $n=2$) is
\begin{eqnarray}
&&a_{2\tau}=-(\widetilde{\vu}\cdot\nabla)\,a_1-
a_{0s}\label{2-appr-1}
\end{eqnarray}
The use of (\ref{A01-appr-1a}) and $\overline{a}_{2\tau}\equiv 0$ transforms (\ref{2-appr-1}) into
\begin{eqnarray}
&&\widetilde{a}_{2\tau}=-(\widetilde{\vu}\cdot\nabla)\,\overline{a}_1- (\widetilde{\vu}\cdot\nabla)\,
\widetilde{a}_1-
\overline{a}_{0s}\label{2-appr-2}
\end{eqnarray}
Its bar-part is
\begin{eqnarray}
&&\overline{a}_{0s}=-\langle(\widetilde{\vu}\cdot\nabla)\,\widetilde{a}_1\rangle
\label{2-appr-4}
\end{eqnarray}
where we have used $\langle{\widetilde{a}_{2\tau}}\rangle=0$,
$\langle(\widetilde{\vu}\cdot\nabla)\,\overline{a}_1\rangle=0$, and $\langle
\overline{a}_{0s}\rangle= \overline{a}_{0s}$.
The substitution of (\ref{01-appr-7}) into (\ref{2-appr-4}) produces the equation
\begin{eqnarray}\label{2-appr-5}
&&\overline{a}_{0s}=
\langle(\widetilde{\vu}\cdot\nabla)(\widetilde{\vxi}\cdot\nabla)\rangle\,\overline{a}_0
\end{eqnarray}
One may expect that the right hand side of (\ref{2-appr-5}) contains both the first and the second spatial derivatives of $\overline{a}_0$, however \emph{all the second
derivatives vanish}. In order to prove it, we introduce the commutator
\begin{eqnarray}
&&\vK\equiv[\widetilde{\vxi},\widetilde{\vu}]=
(\widetilde{\vu}\cdot\nabla)\widetilde{\vxi}-(\widetilde{\vxi}\cdot\nabla)\widetilde{\vu},\label{App1-2}\\
&&\vK\cdot\nabla=(\widetilde{\vu}\cdot\nabla)(\widetilde{\vxi}\cdot\nabla)-(\widetilde{\vxi}\cdot\nabla)(\widetilde{\vu}\cdot\nabla)
\label{App1-1}
\end{eqnarray}
The bar-part of (\ref{App1-1}) is
\begin{eqnarray}
&&\langle(\widetilde{\vu}\cdot\nabla)(\widetilde{\vxi}\cdot\nabla)\rangle
=\langle(\widetilde{\vxi}\cdot\nabla)(\widetilde{\vu}\cdot\nabla)\rangle+\overline{\vK}\cdot\nabla\label{App1-3}
\end{eqnarray}
At the same time, the integration by parts over $\tau$ gives
\begin{eqnarray}
&&\langle(\widetilde{\vu}\cdot\nabla)(\widetilde{\vxi}\cdot\nabla)\rangle
=-\langle(\widetilde{\vxi}\cdot\nabla)(\widetilde{\vu}\cdot\nabla)\rangle,\quad\widetilde{\vu}\equiv\widetilde{\vxi}_\tau
\label{App1-4}
\end{eqnarray}
Combining (\ref{App1-3}) with (\ref{App1-4}) we obtain
\begin{eqnarray}
&&\langle(\widetilde{\vu}\cdot\nabla)(\widetilde{\vxi}\cdot\nabla)\rangle
=\frac{1}{2}\overline{\vK}\cdot\nabla\label{App1-5}
\end{eqnarray}
which reduces (\ref{2-appr-5}) to the advection equation (\ref{4.15}) with
\begin{eqnarray}
&&\overline{\vV}_0\equiv-\langle(\widetilde{\vu}\cdot\nabla)\,\widetilde{\vxi}\rangle=
-\frac{1}{2}\langle[\widetilde{\vxi},\widetilde{\vu}]\rangle=-\frac{1}{2}\overline{\vK}
\label{2-appr-8}
\end{eqnarray}
which also gives the main term in drift velocity (\ref{4.18}). The tilde-part of (\ref{2-appr-2}) appears after subtracting (\ref{2-appr-4})
from (\ref{2-appr-2}):
\begin{eqnarray}
&&\widetilde{a}_{2\tau}= -(\widetilde{\vu}\cdot\nabla)\, \overline{a}_1 -
\{(\widetilde{\vu}\cdot\nabla)\,\widetilde{a}_1\}.
\label{2-appr-10}
\end{eqnarray}
Its tilde-integration with the use of (\ref{01-appr-7}) gives (\ref{A4.12}):
\begin{eqnarray}
&&\widetilde{a}_{2}=-(\widetilde{\vxi}\cdot\nabla)\, \overline{a}_1 +
\{(\widetilde{\vu}\cdot\nabla)(\widetilde{\vxi}\cdot\nabla)\}^\tau\overline{a}_0
\label{2-appr-11}
\end{eqnarray}
Hence, $a_2$ can be written as
\begin{eqnarray}
&&a_2=\overline{a}_2+\widetilde{a}_2
\quad
\label{2-appr-12}
\end{eqnarray}
where $\overline{a}_0$ and $\widetilde{a}_{2}$ are given by (\ref{4.15}), (\ref{2-appr-11}), while $\overline{a}_1,\overline{a}_2$ are not defined.

\underline{\emph{The third-order equation}} ((\ref{Abasic-7}) for $n=3$) is:
\begin{eqnarray}
&&\widetilde{a}_{3\tau}=-(\widetilde{\vu}\cdot\nabla)\, a_2-a_{1s}\label{3-appr-1}
\end{eqnarray}
Its bar-part is
\begin{eqnarray}
&&\overline{a}_{1s}= -\langle(\widetilde{\vu}\cdot\nabla)\,\widetilde{a}_2\rangle.
\label{3-appr-2}
\end{eqnarray}
The substitution of (\ref{2-appr-11}) into (\ref{3-appr-2}), the use of  $\widetilde{\vu}\equiv\widetilde{\vxi}_\tau$,
and the integration by parts yield
\begin{eqnarray}
&&\overline{a}_{1s}=
\langle(\widetilde{\vu}\cdot\nabla)(\widetilde{\vxi}\cdot\nabla)\rangle\overline{a}_1+
\langle(\widetilde{\vxi}\cdot\nabla)(\widetilde{\vu}\cdot\nabla)(\widetilde{\vxi}\cdot\nabla)\rangle\overline{a}_0
\label{3-appr-3}
\end{eqnarray}
where $\langle(\widetilde{\vu}\cdot\nabla)(\widetilde{\vxi}\cdot\nabla)\rangle$ has been already simplified in
(\ref{App1-5}). The second term in right hand side of (\ref{3-appr-3}) formally contains the third, the second, and the first spatial
derivatives of $\overline{a}_0$; however \emph{all the third  and  the second derivatives vanish}. To prove it, first, we
use (\ref{oper-9a}):
\begin{eqnarray}
&&\langle(\widetilde{\vxi}\cdot\nabla)(\widetilde{\vu}\cdot\nabla)(\widetilde{\vxi}\cdot\nabla)\rangle=
-\langle(\widetilde{\vu}\cdot\nabla)(\widetilde{\vxi}\cdot\nabla)(\widetilde{\vxi}\cdot\nabla)\rangle-
\langle(\widetilde{\vxi}\cdot\nabla)(\widetilde{\vxi}\cdot\nabla)(\widetilde{\vu}\cdot\nabla)\rangle\label{App2-1}
\end{eqnarray}
Then we use (\ref{App1-2}), (\ref{App1-1}) to transform the sequence of operators $(\widetilde{\vxi}\cdot\nabla)$ and
$(\widetilde{\vu}\cdot\nabla)$ in each term in right hand side of (\ref{App2-1}) into their sequence in left hand side. The result is
\begin{eqnarray}
&&\langle(\widetilde{\vxi}\cdot\nabla)(\widetilde{\vu}\cdot\nabla)(\widetilde{\vxi}\cdot\nabla)\rangle=
\frac{1}{3}\overline{\vK'}\cdot\nabla,\quad \vK'\equiv[\vK,\widetilde{\vxi}]\label{App2-2}
\end{eqnarray}
As the result (\ref{3-appr-3}) takes a form (\ref{4.16})  with $\overline{\vV}_0$ (\ref{2-appr-8}) and
\begin{eqnarray}
&&\overline{\vV}_1\equiv-\langle(\widetilde{\vxi}\cdot\nabla)(\widetilde{\vu}\cdot\nabla)\widetilde{\vxi})\rangle=
-\frac{1}{3}\langle[[\widetilde{\vxi},\widetilde{\vu}],\widetilde{\vxi}]\rangle=-\frac{1}{3}\overline{\vK'}
\label{3-appr-4a}
\end{eqnarray}
which gives (\ref{4.18}). The tilde-part of (\ref{3-appr-1}) after its integration gives
\begin{eqnarray}
&&\widetilde{a}_{3}=-(\widetilde{\vxi}\cdot\nabla)\,\overline{a}_2-
\{(\widetilde{\vu}\cdot\nabla)\,\widetilde{a}_2\}^\tau-\widetilde{a}_{1s}^\tau,
\quad\widetilde{a}_1^\tau=-(\widetilde{\vxi}^\tau\cdot\nabla)\, \overline{a}_0 \label{3-appr-5}
\end{eqnarray}
where $\widetilde{a}_2$ is given by (\ref{2-appr-11}).
Hence, $a_3$ can be written as
\begin{eqnarray}
&&a_3=\overline{a}_3+\widetilde{a}_3
\label{3-appr-6}
\end{eqnarray}
where $\overline{a}_0$, $\overline{a}_1$, $\widetilde{a}_{2}$, and $\widetilde{a}_{3}$ are given by (\ref{4.15}),
(\ref{4.16}), (\ref{2-appr-11}), and (\ref{3-appr-5}), while $\overline{a}_2,\overline{a}_3$ are not defined.

\underline{\emph{The fourth-order equation}} ((\ref{Abasic-7}) for $n=4$) is:
\begin{eqnarray}
&&\widetilde{a}_{4\tau}=-(\widetilde{\vu}\cdot\nabla)\, a_3-a_{2s}\label{4-appr-1}
\end{eqnarray}
Its bar-part is
\begin{eqnarray}
&&\overline{a}_{2s}= -\langle(\widetilde{\vu}\cdot\nabla)\,\widetilde{a}_3\rangle
\label{4-appr-2}
\end{eqnarray}
The substitution of $\widetilde{a}_3$ (\ref{3-appr-5})  into (\ref{4-appr-2}), $\widetilde{a}_2$ (\ref{2-appr-11}) into
$\widetilde{a}_3$ (\ref{3-appr-5}), the integration by parts (\ref{oper-9}), and the use of
(\ref{2-appr-8}), (\ref{3-appr-4a}) yield
\begin{eqnarray}
&&\langle(\widetilde{\vu}\cdot\nabla)\,\widetilde{a}_3\rangle=(\overline{\vV}_0\cdot\nabla)\overline{a}_2+
(\overline{\vV}_1\cdot\nabla)\overline{a}_1+\langle(\widetilde{\vxi}\cdot\nabla)(\widetilde{\vxi}\cdot\nabla)\rangle
(\overline{\vV}_0\cdot\nabla)\overline{a}_0-\label{App3-1}\\
&&-\langle(\widetilde{\vxi}\cdot\nabla)(\widetilde{\vxi}_s\cdot\nabla)\rangle\overline{a}_0+
\overline{\mathfrak{X}}\overline{a}_0,\ \text{where}\
\overline{\mathfrak{X}}\equiv\langle(\widetilde{\vxi}\cdot\nabla)(\widetilde{\vu}\cdot\nabla)
\{(\widetilde{\vu}\cdot\nabla)(\widetilde{\vxi}\cdot\nabla)\}^\tau\rangle
\nonumber
\end{eqnarray}
The `Gothic' shorthand operator $\overline{\mathfrak{X}}$ (as well as operators $\mathfrak{Y}$,
$\mathfrak{\overline{A}}$, $\mathfrak{\overline{B}}$, $\overline{\mathfrak{C}}$, and $\overline{\mathfrak{F}}$ below)
acts on $\overline{a}_0$. Right hand side of (\ref{App3-1}) formally contains the fourth, the third, the second, and the first spatial
derivatives of $\overline{a}_0$; however \emph{all the fourth and the third derivatives vanish}. In order to prove it we
first rewrite $\overline{\mathfrak{X}}$ as
\begin{eqnarray}
&&\overline{\mathfrak{X}}=\langle(\widetilde{\vxi}\cdot\nabla)(\widetilde{\vu}\cdot\nabla)
\widetilde{\mathfrak{Y}}^\tau\rangle\quad\text{where}\quad
\mathfrak{Y}\equiv(\widetilde{\vu}\cdot\nabla)(\widetilde{\vxi}\cdot\nabla)
\label{App3-2}
\end{eqnarray}
The use of (\ref{oper-9a}) and (\ref{App1-2}), (\ref{App1-1}) transforms (\ref{App3-2}) into
\begin{eqnarray}
&&2\overline{\mathfrak{X}}=-\overline{\mathfrak{A}}+\overline{\mathfrak{B}}+
\frac{1}{2}\langle(\widetilde{\vxi}\cdot\nabla)(\widetilde{\vxi}\cdot\nabla)\rangle(\overline{\vK}\cdot\nabla)
\label{App3-3}\\
&&\overline{\mathfrak{A}}\equiv\langle(\widetilde{\vxi}\cdot\nabla)(\widetilde{\vxi}\cdot\nabla)
(\widetilde{\vu}\cdot\nabla)(\widetilde{\vxi}\cdot\nabla)\rangle,\quad
\overline{\mathfrak{B}}\equiv\langle(\widetilde{\vK}^\tau\cdot\nabla)(\widetilde{\vu}\cdot\nabla)
(\widetilde{\vxi}\cdot\nabla)\rangle
\nonumber
\end{eqnarray}
Let us now simplify $\overline{\mathfrak{A}}$ and $\overline{\mathfrak{B}}$.  For $\overline{\mathfrak{B}}$ we use
(\ref{oper-9a})
\begin{eqnarray}
&&\overline{\mathfrak{B}}\equiv
\langle(\widetilde{\vK}^\tau\cdot\nabla)(\widetilde{\vu}\cdot\nabla)(\widetilde{\vxi}\cdot\nabla)\rangle=-
\langle(\widetilde{\vK}\cdot\nabla)(\widetilde{\vxi}\cdot\nabla)(\widetilde{\vxi}\cdot\nabla)\rangle-
\langle(\widetilde{\vK}^\tau\cdot\nabla)(\widetilde{\vxi}\cdot\nabla)(\widetilde{\vu}\cdot\nabla)\rangle\nonumber
\end{eqnarray}
To change $(\widetilde{\vxi}\cdot\nabla)(\widetilde{\vu}\cdot\nabla)$ into
$(\widetilde{\vu}\cdot\nabla)(\widetilde{\vxi}\cdot\nabla)$ in the last term we use (\ref{App1-2}), (\ref{App1-1}) that
yields:
\begin{eqnarray}
&&\overline{\mathfrak{B}}=-\frac{1}{2}\langle(\widetilde{\vK}\cdot\nabla)(\widetilde{\vxi}\cdot\nabla)
(\widetilde{\vxi}\cdot\nabla)\rangle-\frac{1}{4}\overline{\vkappa}\cdot\nabla,\quad
\vkappa\equiv[\widetilde{\vK}^\tau,\widetilde{\vK}]\label{App3-4}
\end{eqnarray}
The operator $\overline{\mathfrak{A}}$ is simplified by the version of (\ref{oper-9a}) with four multipliers
\begin{eqnarray}
&&\mathfrak{\overline{A}}\equiv\langle(\widetilde{\vxi}\cdot\nabla)(\widetilde{\vxi}\cdot\nabla)
(\widetilde{\vu}\cdot\nabla)(\widetilde{\vxi}\cdot\nabla)\rangle=
-\langle(\widetilde{\vu}\cdot\nabla)(\widetilde{\vxi}\cdot\nabla)(\widetilde{\vxi}\cdot\nabla)
(\widetilde{\vxi}\cdot\nabla)\rangle-\label{App3-5}\\
&&-\langle(\widetilde{\vxi}\cdot\nabla)(\widetilde{\vu}\cdot\nabla)(\widetilde{\vxi}\cdot\nabla)
(\widetilde{\vxi}\cdot\nabla)\rangle-
\langle(\widetilde{\vxi}\cdot\nabla)(\widetilde{\vxi}\cdot\nabla)
(\widetilde{\vxi}\cdot\nabla)(\widetilde{\vu}\cdot\nabla)\rangle
\nonumber
\end{eqnarray}
The multiple use of commutator (\ref{App1-2}), (\ref{App1-1})  allows us to transform the sequence of operators
$(\widetilde{\vxi}\cdot\nabla)$ and $(\widetilde{\vu}\cdot\nabla)$ in each term in right hand side of (\ref{App3-5}) to the
sequence in its LHS. The result is
\begin{eqnarray}
&&\overline{\mathfrak{A}}=-\frac{1}{2}\langle(\vK\cdot\nabla)(\widetilde{\vxi}\cdot\nabla)
(\widetilde{\vxi}\cdot\nabla)\rangle+\frac{1}{4}\overline{\vK''}\cdot\nabla,\quad
\vK''\equiv[\vK',\widetilde{\vxi}]\label{App3-6}
\end{eqnarray}
Now, (\ref{App3-3}), (\ref{App3-4}), and (\ref{App3-6}) yield
\begin{eqnarray}
&&\overline{\mathfrak{X}}=
\frac{1}{4}(\overline{\vK}\cdot\nabla)\langle(\widetilde{\vxi}\cdot\nabla)(\widetilde{\vxi}\cdot\nabla)\rangle+
\frac{1}{4}\langle(\widetilde{\vxi}\cdot\nabla)(\widetilde{\vxi}\cdot\nabla)\rangle(\overline{\vK}\cdot\nabla)-
\frac{1}{8}(\overline{\vkappa}+\overline{\vK''})\cdot\nabla\nonumber
\label{App3-7}
\end{eqnarray}
The substitution of this expression into (\ref{App3-1}), (\ref{4-appr-2})  gives
\begin{eqnarray}
&&\overline{a}_{2s}+(\overline{\vV}_0\cdot\nabla)\overline{a}_2+
(\overline{\vV}_1\cdot\nabla)\overline{a}_1-\frac{1}{8}(\overline{\vkappa}+\overline{\vK''})\cdot\nabla \overline{a}_0
+\frac{1}{4}\overline{\mathfrak{C}}\overline{a}_0-\overline{\mathfrak{F}}\overline{a}_0=0,\label{App3-8}\\
&&\overline{\mathfrak{C}}\equiv(\overline{\vK}\cdot\nabla)\langle(\widetilde{\vxi}\cdot\nabla)(\widetilde{\vxi}\cdot\nabla)
\rangle-\langle(\widetilde{\vxi}\cdot\nabla)(\widetilde{\vxi}\cdot\nabla)\rangle(\overline{\vK}\cdot\nabla),\
\overline{\mathfrak{F}}\equiv\langle(\widetilde{\vxi}\cdot\nabla)(\widetilde{\vxi}_s\cdot\nabla)\rangle
\nonumber
\end{eqnarray}
Additional transformations of the last two operators in (\ref{App3-8}) yield
\begin{eqnarray}
&&\frac{1}{4}\overline{\mathfrak{C}}-\overline{\mathfrak{F}}=
\frac{1}{2}\langle[\widetilde{\vxi},\widetilde{\vxi}_s]\rangle\cdot\nabla
-\frac{1}{2}\langle(\widetilde{\vu}'\cdot\nabla)\widetilde{\vxi}+
(\widetilde{\vxi}\cdot\nabla)\widetilde{\vu}'\rangle\cdot\nabla - \frac{1}{2}\langle
\widetilde{u}'_i\widetilde{\xi}_k+\widetilde{u}'_k\widetilde{\xi}_i\rangle\frac{\partial^2}{\partial x_i\partial x_k}=\nonumber\\
&&=\frac{1}{2}\langle[\widetilde{\vxi},\widetilde{\vxi}_s]\rangle\cdot\nabla-\frac{\partial}{\partial
x_k}\left(\overline{\chi}_{ik}\frac{\partial}{\partial
x_i}\right)+\frac{1}{2}\langle\widetilde{\vxi}\Div\widetilde{\vu}'+\widetilde{\vu}'\Div\widetilde{\vxi}\rangle
\label{Ap-trans}\\
&&\widetilde{\vu}'\equiv\widetilde{\vxi}_s-[\overline{\vV}_0,\widetilde{\vxi}],\quad
\overline{\chi}_{ik}\equiv\frac{1}{2}\langle \widetilde{u}'_i\widetilde{\xi}_k+\widetilde{u}'_k\widetilde{\xi}_i\rangle
 \label{v-prime}
\end{eqnarray}
The  substitution of (\ref{Ap-trans}) into (\ref{App3-8}) leads to the equation for $\overline{a}_2$ (\ref{4.17}) where
the formula (\ref{4.20}) for $\overline{\chi}_{ik}$ is obtained from (\ref{v-prime}) by the use of definition
$\widetilde{\vu}'$.
The tilde-part of (\ref{4-appr-1}) after its tilde-integration gives (\ref{A4.14})
\begin{eqnarray}
&&\widetilde{a}_{4}=-(\widetilde{\vxi}\cdot\nabla)\,\overline{a}_3-
\{(\widetilde{\vu}\cdot\nabla)\,\widetilde{a}_3\}^\tau-\widetilde{a}_{2s}^\tau \label{4-appr-5}
\end{eqnarray}
where $\widetilde{a}_3$ is given by (\ref{3-appr-5}).
Hence, we have solved the equation \eqref{Abasic-2} for the first five approximations and obtained the correspondent required terms in
(\ref{basic-4aa}).


\begin{thebibliography}

\bibitem[Andrews \& McIntyre (1978)]{McIntyre} \textsc{Andrews, D. and McIntyre, M.E.} 1978
An exact theory of nonlinear waves on a Lagrangian-mean flow. {\it J. Fluid Mech.}, \textbf{89}, 609-646.

\bibitem[Aref (1984)]{Aref} \textsc{Aref, H.} 1984
Stirring by chaotic advection. {\it J. Fluid Mech.}, \textbf{143}, 1-21.

\bibitem[Arnold (1964)]{Arnold} \textsc{Arnold, V.I.} 1964
Instabilities of dynamical systems with several degrees of freedom. {\it Sov. Math. Dokl.}, \textbf{6}, 581-585.

%

\bibitem[Batchelor (1967)]{Batchelor} \textsc{Batchelor, G.K.} 1967 {\it An Introduction to Fluid Dynamics.}
Cambridge: CUP.




%
%
%

\bibitem[Buhler (2009)]{Buhler} \textsc{Buhler, O.} 2009 {\it Waves and Mean Flows.}
Cambridge: CUP.

%
%



\bibitem[Chierchia \& Gallavotti (1994)]{Chierchia} \textsc{Chirchia, L. \&  Gavalotti, G.} 1994
Drift and diffusion in phase space, \emph{Annales de l'I.H.P., section Physique Th$\acute{e}$orique,} \textbf{60},
1-144.


\bibitem[Craik (1982)]{Craik} \textsc{Craik, A.D.D.} 1982 The drift velocity of water waves.
\emph{J. Fluid Mech.}, \textbf{116}, 187-205.

\bibitem[Craik (1985)]{Craik0} \textsc{Craik, A.D.D.} 1985 \emph{Wave Interactions and Fluid Flows.}
Cambridge, CUP.

\bibitem[Darwin (1953)]{Darwin} \textsc{Darwin, C.} 1953 Note on hydrodynamics.
\emph{Proc. Camb. Phil. Soc.}, \textbf{49}, 342-354.

\bibitem[Debnath (1994)]{Debnath} \textsc{Debnath, L.} 1994 {\it Nonlinear Water Waves.}
Boston: Academic Press.


\bibitem[Dombre, Frish, et al. (1986)]{Frish} \textsc{Dombre, T., Frish, U., Greene, J.M., Henon,M, Mehr, A, and Soward, A.M.} 1986
Chaotic streamlines in the ABC flows \emph{J. Fluid Mech.}, \textbf{167}, 353–391.


\bibitem[Eames, Belcher, and Hunt (1994)]{Hunt1} \textsc{Eames, I., Belcher, S.E., and Hunt, J.C.R.} 1994
Drift, partial drift and Darwin's proposition. \emph{J. Fluid Mech.}, \textbf{275}, 201-223.


\bibitem[Eames \& McIntyre (1999)]{Eames} \textsc{Eames, I. \& McIntyre, M.E.} 1999
On the connection between the Stokes drift and Darwin drift.
\emph{Math. Proc. Camb. Phil. Soc.}, \textbf{126}, 171-174.



\bibitem[Grimshaw (1984)]{Grimshaw} \textsc{Grimshaw, R.} 1984 Wave action and wave-mean flow interactions,
with applications to stratified shear flow. \emph{Ann. Rev. Fluid Mech.}, \textbf{16}, 11-44.





\bibitem[Kevorkian \& Cole (1996)]{Kevorkian} \textsc{Kevorkian, J. \& Cole, J.D.} 1996
\emph{Multiple Scale and Singular Perturbation Methods}, NY: Springer.

\bibitem[Lamb (1932)]{Lamb} \textsc{Lamb, H.} 1932 \emph{Hydrodynamics.} Sixth edition,
Cambridge, CUP.

%
%
%

{}\bibitem[Levenshtam (1996)]{Levenshtam} \textsc{Levenshtam, V.B.} 1996
Justification of averaging method for thermal vibrational convection.
\emph{Dokl. Akad. Nauk.}, \textbf{349}, 5, 621-623 (in Russian).


{}\bibitem[Lighthill (1956)]{Lighthill} \textsc{Lighthill, M. J.} 1956 Drift. \emph{J.Fluid Mech.}, \textbf{1},
31-54 (and Corrigendum \textbf{2}, 311-312).


\bibitem[Longuet-Higgins (1953)]{LH} \textsc{Longuet-Higgins, M. S.} 1953 Mass transport in water waves.
\emph{Phil. Trans. A,} \textbf{245}, 535-581.

\bibitem[Maxwell (1870)]{Maxwell} \textsc{Maxwell, J. C.} 1870 On the displacement in a case of fluid motion,
\emph{Proc. London Math. Soc.}, \textbf{3}, 82-87.

\bibitem[McLachlan (1962)]{mathieu} \textsc{McLachlan, N.W.} 1962
\emph{Theory and application of Mathieu functions.} New York: Dover.
Cambridge: CUP.

\bibitem[Mizrahi \& Daboul (1992)]{Mizrahi} \textsc{Mizrahi, S.S. and Daboul, J.} 1992
Squeezed states, generalized Hermite polynomials and pseudo-diffusion equation.
\emph{Physica A}, \textbf{189}, 635-650.

\bibitem[Moffatt (1978)]{Moffatt} \textsc{Moffatt, H.K.} 1978
\emph{Magnetic Field Generation in Electrically Conducting Fluid.}
Cambridge: CUP.

\bibitem[Nayfeh (1973)]{Nayfeh} \textsc{Nayfeh, A.H.} 1973 \emph{Perturbation Methods.} NY: John Wiley \& Sons.

\bibitem[Ottino (1989)]{Ottino} \textsc{Ottino, T.J.} 1989
{\it The Kinematics of Mixing. Stretching, Chaos, and Transport.} Cambridge: CUP.

\bibitem[Samelson \& Wiggins (2006)]{Wiggins} \textsc{Samelson, R.M. \& Wiggins, S.} 2006
\emph{Lagrangian Transport in Geophysical Jets and Waves: the Dynamical Systems Approach}, NY: Springer.

\bibitem[Sanders \& Verhulst (1985)]{Verhulst} \textsc{Sanders, J.A. and Verhulst, F.} 1985 \emph{Averaging
Methods in Nonlinear Dynamical Systems.} \emph{Applied Mathematical Sciences.}  \textbf{59}, NY: Springer.


\bibitem[Simonenko (1972)]{Simonenko} \textsc{Simonenko, I.B.} 1972 Justification of averaging method for convection
problem in the field of rapidly oscillating forces and for other parabolic equations. \emph{Math. Sbornik},
\textbf{87(129)}, 2, 236-253 (in Russian).

%

\bibitem[Stokes (1847)]{Stokes} \textsc{Stokes, G.G.} 1847 On the theory of oscillatory waves.
\emph{Trans. Camb. Phil. Soc.}, \textbf{8}, 441-455 (Reprinted in \emph{Math. Phys. Papers}, \textbf{1}, 197-219).




\bibitem[Vladimirov (2005)]{Vladimirov2} \textsc{Vladimirov, V.A.}
2005  Vibrodynamics of pendulum and submerged solid.  {\it J. of Math. Fluid Mech.} \textbf{7}, S397-412.

\bibitem[Vladimirov (2008)]{Vladimirov1} \textsc{Vladimirov, V.A.}
2008 Viscous flows in a half space caused by tangential vibrations on its boundary. {\it Studies in Appl. Math.},
\textbf{121}, 4, 337-367.


\bibitem[Vladimirov (2010)] {VladimirovDr1}\textsc{Vladimirov, V.A.} 2010
Admixture and drift in oscillating fluid flows.
\emph{ArXiv}: 1009, 4085v1, (physics, flu-dyn).


\bibitem[Vladimirov (2011)] {VladimirovDr2}\textsc{Vladimirov, V.A.} 2011
Theory of non-degenerate oscillatory flows.
\emph{ArXiv}: 1110, 3633v2, (physics, flu-dyn).


\bibitem[Vladimirov (2012)]{VladimirovL} \textsc{Vladimirov, V.A.}
2012 Magnetohydrodynamic drift equations: from Langmuir circulations to magnetohydrodynamic dynamo? {\it J. of Fluid Mech.},
\textbf{698}, 51-61.


\bibitem[Vladimirov, Proctor and Hughes (2015)]{VladProc} \textsc{Vladimirov, V.A., Proctor, M.R.E.  and Hughes, D.W} 2015,
Vortex dynamics of oscillating flows. \emph{Arnold Math J.}, \textbf{239}, 2, 113-126.



\bibitem[Yudovich (2006)]{Yudovich} \textsc{Yudovich, V.I.} 2006
Vibrodynamics and vibrogeometry of mechanical systems with constrains.
\emph{Uspehi Mekhaniki,} \textbf{4}, 3, 26-158 (in Russian).






\end{thebibliography}
\end{document}